\documentclass[a4paper,11pt,table]{article}
\usepackage{amssymb,amsmath,latexsym,color,amsthm,authblk,mathpazo,palatino,bbding,setspace,datetime,breakcites,hyphenat,microtype,diagbox,tikz,xcolor,subcaption,graphicx,titlesec}
\usepackage{mathptmx}
\usepackage[utf8]{inputenc}
\usepackage[round,authoryear]{natbib}
\usepackage{tikz-network}
\usepackage{hyperref,theoremref}
\usepackage[titletoc,title]{appendix}
\usepackage{mathtools}
\usepackage{bm}
\usepackage{esvect}
\usepackage{mathtools}
\usepackage{bm}
\usepackage{esvect}
\usepackage{amsfonts}
\usepackage{mathrsfs}
\usepackage{amssymb}
\usepackage{yfonts}
\usepackage{dsfont}
\usepackage{csquotes}
\usepackage{multicol}
\usepackage{lipsum}
\usepackage{accents}
\usepackage{titlecaps, stringstrings}

\usepackage[shortlabels]{enumitem}
\usepackage{float}
\usepackage{amssymb}
\restylefloat{table}

\definecolor{armygreen}{rgb}{0.29, .8, 0.13}
\definecolor{auburn}{rgb}{0.43,0.21, 0.1}
\definecolor{burgundy}{rgb}{0.5,0.0, 0.13}
\definecolor{medium red}{rgb}{.490,.298,.337}
\definecolor{dark red}{rgb}{.235,.141,.161}

\hypersetup{
	colorlinks = true,
	linkcolor = {burgundy},
	citecolor = {burgundy},
	linkbordercolor = {white},
}

\let\OLDthebibliography\thebibliography
\renewcommand\thebibliography[1]{
	\OLDthebibliography{#1}
	\setlength{\parskip}{0pt}
	\setlength{\itemsep}{0pt plus 0.1ex}
}

\DeclareFontFamily{U}{mathx}{\hyphenchar\font45}
\DeclareFontShape{U}{mathx}{m}{n}{<-> mathx10}{}

\titleformat{\section}[block]{\normalfont\scshape\large\filcenter}{\thesection .}{1em}{}
\titleformat{\subsection}{\normalfont\scshape\large}{\thesubsection}{1em}{}
\titleformat{\subsubsection}{\normalfont\scshape\large}{\thesubsubsection}{1em}{}

\newtheorem{theorem}{Theorem}[section]
\newtheorem{proposition}{Proposition}[section]

\theoremstyle{definition}

\theoremstyle{remark}

\newcommand{\kbar}{\overline{k}}

\newdateformat{monthyeardate}{\monthname[\THEMONTH], \THEYEAR}

\title{\textsc{}\thanks{\noindent}}

\author[1]{Souvik Roy\footnote{Corresponding Author: souvik.2004@gmail.com}}
\author[1]{Agamani Saha\footnote{Contact: agamanisaha@gmail.com}}
\affil[1]{Applied Statistics Unit, Indian Statistical Institute, Kolkata}
\date{}

\title{\textsc{A Directed Lazy Random Walk Model to Three-Way Dynamic Matching Problem}}
\begin{document}
	
	\maketitle
\begin{abstract}
    This paper explores a novel extension of dynamic matching theory by analyzing a three-way matching problem involving agents from three distinct populations, each with two possible types. Unlike traditional static or two-way dynamic models, our setting captures more complex team-formation environments where one agent from each of the three populations must be matched to form a valid team. We consider two preference structures: assortative or homophilic, where agents prefer to be matched with others of the same type, and dis-assortative or heterophilic, where diversity within the team is valued. Agents arrive sequentially and face a trade-off between matching immediately or waiting for a higher quality match in the future albeit with a waiting cost. We construct and analyze the corresponding transition probability matrices for each preference regime and demonstrate the existence and uniqueness of stationary distributions. Our results show that stable and efficient outcomes can arise in dynamic, multi-agent matching environments, offering a deeper understanding of how complex matching processes evolve over time and how they can be effectively managed.
\end{abstract}
\section{Introduction}
\subsection{Description of the problem} Dynamic matching problems in game theory refer to environments where agents enter the market sequentially over time, each seeking to be matched with others based on certain criteria or preferences. These interactions yield payoffs or utilities for the agents involved, depending on the quality and timing of their matches. A central feature of such settings is the inherent trade-off between the potential benefits of waiting for a more desirable match and the costs associated with delay or prolonged participation in the market. The goal is to design matching mechanisms that efficiently balance these trade-offs and maximize the overall welfare generated in society.

In a dynamic context, agents must often make decisions under uncertainty, not only about the preferences of other agents but also about the timing and availability of future arrivals. This complexity distinguishes dynamic matching from static settings and introduces new challenges in mechanism design and equilibrium analysis.

While much of the prior work in this area has focused on matching between two populations with a limited number of types, this paper extends the framework to a more general setting involving three distinct populations. Each population consists of agents belonging to one of two possible types. A valid matching consists of forming a team with one agent from each population. We explore different scenarios based on type preferences, one where agents prefer matching with others of the same type and another where diversity of types within a team is preferred.

\subsection{Motivation of the problem}
This paper extends the concept of two-way matching to the situation where there are three populations, each consisting of two types of agents. This may be referred to as a three-way matching problem. Such a framework captures more complex interactions compared to traditional two-sided markets and has practical relevance in several real-world scenarios.

For instance, in the healthcare domain, teams are often composed of a doctor, a nurse, and a technician. Each professional may be junior or experienced, and hospitals aim to assemble teams that provide a balanced mix of expertise to ensure quality care. Consider another example where participants from three regions, say US, EU and Asia are characterized as doves or hawks, representing aggressive or conciliatory negotiation styles. A team consisting one participant from each region need to be formed. Teams are often designed to ensure ideological diversity to simulate realistic negotiations. Similarly, in corporate product development, teams typically consist of an engineer, a designer, and a product manager, with each individual leaning toward either an agile or traditional mindset. The productivity and adaptability of such teams often hinge on achieving the right mix of work styles.

\subsection{Related Literature}

The classical literature on matching theory has largely focused on static, two-sided environments where all agents are present simultaneously, and matchings occur between individuals from two distinct populations. Foundational contributions such as \citet{gale1962college} introduced the notion of stable matchings via the deferred acceptance algorithm, providing a basis for many centralized allocation mechanisms. \citet{becker1974theory} offered a structural economic model of marriage, emphasizing sorting and preference-based match formation. Subsequent empirical work, including \citet{hitsch2010matching}, explored these theoretical insights using online dating data, identifying patterns of assortative matching and strategic behavior.

While static models offer clean characterizations, many real-world applications involve dynamic settings where agents arrive and depart over time. This dynamic perspective introduces intertemporal trade-offs, coordination challenges, and uncertainty regarding future market participants. Notable work in this area includes \citet{zenios1999modeling}, who employed a queueing model with reneging to study transplant waiting lists, and \citet{unver2010dynamic}, who examined multi-way kidney exchanges and demonstrated that it may be optimal to postpone simple matches to enable more beneficial future ones. These contributions were primarily centered around compatibility rather than preference-driven dynamics.

Building on these foundations, a growing body of research has focused on optimal matching mechanisms in dynamic environments with preferences. \citet{baccara2021optimal} model dynamic matching among two populations with two types in each, where the planner aims to maximize societal welfare under arrival uncertainty. \citet{akbarpour2020thickness} explore a setting where agents may perish if unmatched by a critical time, highlighting how increasing market thickness can mitigate loss due to agent departure. Further developments by \citet{loertscher2018optimal} and \citet{doval2019decentralized} incorporate discounting and strategic delays into welfare analyses. These models remain largely limited to two-way matching between two sides of the market.

However, many real-world allocation problems involve more complex team formation or group matching, going beyond simple pairwise structures. \citet{echenique2015solution} consider environments in which agents form teams and express preferences over their colleagues, representing a departure from the standard two-sided framework. Similarly, \citet{ostrovsky2008stability} examines multi-tiered supply chains, where the stability of multi-level matching becomes critical—an indirect analogue to three-sided markets. Applications such as adoption, labor, or research collaboration often involve forming teams composed of multiple, heterogeneous agents. \citet{baccara2014child} study adoption matches with preferences over both gender and race, underscoring the multidimensional nature of matching.

The organ donation literature also motivates extensions to multi-agent matchings. \citet{roth2005pairwise} develop models of kidney exchange based on pairwise trades, while later work extends this to dynamic multi-way settings, as seen in \citet{unver2010dynamic}. Additionally, school-choice mechanisms like the Boston mechanism studied by \citet{kojima2014boston} reveal how preferences and timing affect centralized allocation in dynamic arrival models.

Decentralized dynamics have also been explored by \citet{haeringer2011decentralized}, where job seekers and firms interact over time without central coordination, introducing frictions that affect efficiency. Queueing models such as those in \citet{naor1969regulation} and \citet{hassin1985excessive} provide further insight into how delay and order-of-service impact matching outcomes, albeit in simplified one-sided environments. \citet{leshno2019dynamic} and \citet{anderson2017finding} analyze online dynamic policies and overloaded systems, contributing algorithms for efficient matching under stochastic agent flows.

\subsection{Contribution of the paper}
In this paper, we extend the dynamic matching framework to a setting involving three distinct populations, where each matching team comprises one individual from each population, and each individual belongs to one of two possible types. We examine two preference structures: one where teams composed of similar types are favored (homophilic preference), and another where diversity in types within a team is preferred (heterophilic perefernce). For both scenarios, we derive the stationary distributions of the corresponding transition probability matrices and show that these distributions are unique. Our results demonstrate that stable outcomes can emerge even in dynamic three-way matching environments, highlighting the tractability and robustness of equilibrium solutions in more complex, multi-agent settings.

\section{Two Way Matching}
\cite{baccara2020optimal} in their paper have studied an infinite-horizon dynamic matching problem where there are 2 kinds of populations $\mathcal{P}_1$ and $\mathcal{P}_2$. Each population $\mathcal{P}_i$, consists of two types of individuals -\enquote*{High} and \enquote*{Low} with probabilities $p_i$ and $1-p_i$ respectively where $i=1,2$. The \enquote*{High} and \enquote*{Low} individuals belonging to population $\mathcal{P}_1$ are denoted as $H$ and $L$ respectively while those belonging to population $\mathcal{P}_2$ are denoted as $h$ and $l$ respectively. $T=\mathbb{N} \cup \{0\}$ denote the set of time points. At each time point $t \in T$, one individual from $\mathcal{P}_1$ and another from $\mathcal{P}_2$ enter the market.

$U_x(y) \in \mathbb{R^+}$ is defined as the utility for a type-x participant belonging to $\mathcal{P}_1$ from matching with a type-y participant belonging to $\mathcal{P}_2$; $x \in \{H,L\}, y \in \{h,l\}$. Here preferences are assumed to be assortative, that is, $H$ type is more desirable for all individuals belonging to $\mathcal{P}_2$ and $h$ type is more desirable for all individuals belonging to $\mathcal{P}_1$. In terms of utilities,
\begin{align*}
    U_H(h) > U_H(l), && U_L(h) > U_L(l),\\
    U_h(H) > U_h(L), && U_l(H) > U_l(L).
\end{align*}
Let
\begin{align*}
    U_{xy} & \equiv U_x(y) + U_y(x) && \forall x,y \in {H,L,h,l}\\
    U & \equiv U_{Hh} + U_{Ll} - U_{Hl} - U_{Lh}.
\end{align*}
It is assumed that $U > 0$, so that utilitarian-efficient matching in a static market creates maximal number of $(H,h)$ and $(L,l)$ pairs. Here $0 < U \leq U_{Hh}$. Each individual suffers a cost $c > 0$ for each time period it spends in the market waiting to be matched. It is further assumed that individuals can leave the market only by matching.
\subsection{Matching Process}
At each time point $t \in \{1,2, ... , \}$, after a new pair has arrived in the market a queue corresponds to 
$s^t = (s_H, s_L, s_h, s_l)$ where each entry is stock of a particular type waiting in line. 
Let $m^t = (m_{Hh}, m_{Hl}, m_{Lh}, m_{Ll})$ be the profile of matches created at time $t$, and 
$k^t = (k_H, k_L, k_h, k_l)$ is the volume of remaining individuals after all matches. 
Initially, $k^0 = (0,0,0,0)$.\\
The utility generated by matches is
\[
S(m) \equiv \sum_{x,y \in \{H,L\} \times \{h, l\}} m_{xy}U{xy}.
\]
The total waiting cost incurred by remaining individuals in period $t$ is
\[
C(s,m) \equiv c\left( \sum_{x \in \{H,L,h,l\}} k_x\right).
\]
The welfare generated at time $t$ is 
\begin{align*}
    w(s,m) & = S(m) - C(s,m) && \text{if profile of matches m is feasible given stock s}\\
    & = -\infty && \text{otherwise.}
\end{align*}

\subsection{Optimal Dynamic Mechanism}
\cite{baccara2020optimal} discusses the case where after a certain waiting time, pairs are matched forcefully irrespective of the their type. The following proposition characterizes this matching problem. 
\begin{proposition} (Optimal mechanisms) An optimal dynamic mechanism is identified by a pair of thresholds $(\overline{k}_{H},\overline{k}_{h}) \in \mathds{Z}_+$ such that:
    \begin{enumerate}[(i)]
        \item If the number of $H$ types present is more than $\overline{k}_{H}$, excess $H$-types are matched immediately, and,
        \item If the number of $h$-types present is more than $\overline{k}_{h}$, excess $h$-types are matched immediately.
    \end{enumerate}
\end{proposition} 
\[
\overline{k} = \overline{k}_{H} = \overline{k}_{h}
\hspace{5mm} \text{ : symmetric thresholds}
\]
$k^t_{Hh}= k^t_H - k^t_h$: signed length of the $H-h$ queue at time point $t$.
\begin{align*}
    k^{t+1}_{Hh} & = k^{t}_{Hh} && \text{if $(H,h)$ or $(L,l)$ appears}\\
    & = k^{t}_{Hh} = \overline{k} && \text{if $(H,l)$ appears and $k^{t}_{Hh} = \overline{k}$}\\
    & = k^{t}_{Hh} = -\overline{k} && \text{if $(L,h)$ appears and $k^{t}_{Hh} = -\overline{k}$}\\
    & = k^{t}_{Hh} + 1 && \text{if $(H,l)$ appears and $-\overline{k} \leq k^{t}_{Hh} < \overline{k}$}\\
    & = k^{t}_{Hh} - 1 && \text{if $(L,h)$ appears and $-\overline{k} < k^{t}_{Hh} \leq\overline{k}$} \\
\end{align*}
This process induces a Markov chain in the following manner. Let $
x^t_i = I_{[k^{t}_{Hh} = i]}$. 
Then, $
x^{t+1} = T_{\overline{k}} x^t$, 
where $T_{\overline{k}}$ is the transition matrix given by 
\[
T_{\overline{k}} = \begin{bmatrix} 
    1-p(1-p) & p(1-p) & \dots  & 0 & 0\\
    p(1-p) & 1-2p(1-p) & \dots  & 0 & 0\\
    0 & p(1-p) & \dots  & 0 & 0\\
    \vdots & \vdots & \vdots & \vdots & \vdots\\
    0 & 0 & \dots  & p(1-p) & 0 \\
    0 & 0 & \dots  & 1-2p(1-p) & p(1-p) \\
    0 & 0 & \dots  & p(1-p) & 1-p(1-p)
    \end{bmatrix}
\]
 It can be shown that this Markov chain is aperiodic, irreducible and positively recurrent. So it is an ergodic Markov chain. Hence, an unique stationary distribution exists. This stationary distribution is derived to be \textit{Uniform}$(-\overline{k}, \overline{k})$.

\section{Three-Way Assortative Matching}
We have extended this above problem to the situation where there are 3 populations $\mathcal{P}_1, \mathcal{P}_2$ and $\mathcal{P}_3$. Each population has two types of individuals - 'High' and 'Low' in the proportion of $p$ and $q = 1-p$ respectively. We make a team of 3 taking one individual from each population. Preference is to take as many high types in a team as possible. Let $H_i$ denote the high type and $L_i$ the low type belonging to population $\mathcal{P}_i$, where $i=1,2,3$. We assume at each time point $t=0,1,2,..$, a triplet, that is, three individuals one from each population, arrive together in the market in the hope of getting a match. \\
We define by $U_x(y,z) \in \mathbb{R^+}$, the utility for a type-x participant belonging to $\mathcal{P}_1$ from matching with a type-y participant belonging to $\mathcal{P}_2$, and a type-z participant belonging to $\mathcal{P}_3$. We assume preferences to be assortative, that is, High type is more desirable for all individuals.\\
That is,
\begin{align*}
    U_{x}(H_j,H_k) >  U_{x}(H_j,L_k) >  U_{x}(L_j,L_k) && \forall x \in \{H_i,L_i\} \text{ and } i,j,k=1,2,3
\end{align*}

\subsection{Matching Process}
Let $k^t = (k_{H_1},k_{L_1},k_{H_2},k_{L_2},k_{H_3},k_{L_3})$ be the volume of remaining individuals at time $t$ after all matches. Let $k^t_H = (k_{H_1},k_{H_2},k_{H_3})$ be the volume of remaining individuals of high type at time $t$.\\
If number of High types belonging to any population exceeds $\kbar$, excess High types are matched immediately. Matching is done in such a way that teams get as many high type of individuals together as possible. In other words, if at time $t$, $k^t_H = \{ \overline{k}+1, a, 0\}, a<\overline{k} $, then a team of $\{H_1,H_2,L_3\}$ will be formed.\\
There are eight possible combinations of incoming individuals as given below: 
\begin{multicols}{3}
    \begin{itemize}[\null]
        \item $H_1H_2L_3$ 
        \item $H_1L_2H_3$ 
        \item $L_1H_2H_3$
        \item $H_1L_2L_3$
        \item $L_1H_2L_3$
        \item $L_1L_2H_3$
        \item $H_1H_2H_3$ 
        \item $L_1L_2L_3$ 
    \end{itemize} 
\end{multicols}
Now, if, $k^t_H = (a_1,a_2,a_3)$, at least one of the $a_i$'s must be 0. Without loss of generality, let us assume $a_3 = 0$.\\
Let us assume that the at time point $t$, there are $a_1$ number of $\mathcal{P}_1$ individuals of high type and  $a_2$ number of $\mathcal{P}_2$ individuals of high type remaining in the market waiting to be matched, i.e., $k^{t}_{H} = (a_1,a_2,a_3) 1 \leq a_1,a_2 \leq \kbar-1$. Then $k^{t+1}_{H}$ takes the value $(a_1+1,a_2,a_3)$ if and only if the combination $\{H_1,L_2,L_3\}$ appears in the next time point, the probability of which is $pq^2$. On the other hand $k^{t}_{H}$ takes the value $(a_1-1,a_2,a_3)$  if and only if $\{H1,L_2,H_3\}$\\ appears and a team consisting of high types is formed. If $a_1=\kbar,a_2$ same as before, then $k^{t}_{H}$ takes the value $(a_1,a_2-1,a_3)$  if ${H_1,L_2,H_3}$ appears and a team of high types is formed or if ${H_1,L_2,L_3}$ and a team ${H_1,H_2,L_3}$ is forcefully formed since $H_1$ is crossing the threshold. This transition has probability $p^q+pq^2=pq$. We can find the probabilities of the other transitions in similar manner. The transition probabilities can be written in the following manner:
\begin{align*}
        k^{t+1}_{H} & = (a_1+1,a_2,a_3), && (a_1+1) \leq \kbar, &&& \text{w.p. } pq^2\\
        & = (a_1-1,a_2,a_3), && a_1,a_2 \neq \kbar, (a_1 - 1) \geq 0 &&& \text{w.p. } p^2q\\
        & = (a_1+1,a_2+1,a_3), && (a_1 + 1) \leq \kbar \text{ and/ or } (a_2 + 1)\leq \kbar, &&& \text{w.p. } p^2q\\
        & = (a_1-1,a_2-1,a_3), && a1,a_2 \neq \kbar,(a_1 - 1) \geq 0 \text{ and/ or } (a_2 - 1)\geq 0, &&& \text{w.p. } pq^2\\
        & = (a_1-1,a_2,a_3), && a_1 = a_2 = \kbar, &&& \text{w.p. } pq\\
        & = (a_1,a_2-1,a_3), && a_1 = \kbar, \quad a_2 = 1,2,..,\kbar-1, &&& \text{w.p. } pq\\
        & = (a_1-1,a_2-1,a_3), && a_1 = a_2 = \kbar, &&& \text{w.p. } pq^2\\
        & = (a_1-1,a_2,a_3), && a_1 = \kbar, \quad a_2 = 0,1,2,..,\kbar-1, &&& \text{w.p. } p^2q\\
        & = (a_1-1,a_2-1,a_3), && a_1 = \kbar, \quad a_2 = 1,2,..,\kbar-1, &&& \text{w.p. } pq^2\\
\end{align*}

Same probabilities hold for similar change in other $a_i$'s, $i=1,2,3$. In other words, the transition probabilities are independent of the order of $a_i$'s, since the proportion of High type is same in all the three populations.\bigskip \\
This process is equivalent to a directed lazy random walk with weighted probabilities on three surfaces of a cube of length $\kbar$. The three surfaces are those which contain the origin point $(0,0,0)$. The person walking on the three surfaces of the cube can move vertically, horizontally or diagonally, only one square at a time, according to the probabilities given above. But diagonal movement is only possible on those lines where values of both axes are increasing or both are decreasing. This is illustrated in the diagram below. The grid is drawn only one surface for clarity. Movement on the other two surfaces will be in similar fashion. \bigskip\\
\begin{tikzpicture}[line cap=round, line join=round, x={(1cm,0cm)}, y={(0cm,1cm)}, z={(-0.6cm,-0.5cm)}]
\draw [->] (0,0,0) -- (4,0,0) node [right] {$Y$};
\draw [->] (0,0,0) -- (0,4,0) node [left]  {$Z$};
\draw [->] (0,0,0) -- (0,0,4) node [left]  {$X$};
\draw[canvas is xz plane at y=0,fill=gray!50] (0,0) rectangle (3,3);
\draw[canvas is yz plane at x=0,fill=gray!50] (0,0) rectangle (3,3);
\begin{scope}[canvas is xy plane at z=0]
\draw(0,0) grid (3,3);
\end{scope}

\draw[->] (0,0,0) -- (1,1,0);
\draw[->] (1,0,0) -- (2,1,0);
\draw[->] (0,1,0) -- (1,2,0);
\draw[->] (1,1,0) -- (2,2,0);
\draw[->] (2,0,0) -- (3,1,0);
\draw[->] (0,2,0) -- (1,3,0);
\draw[->] (2,1,0) -- (3,2,0);
\draw[->] (1,2,0) -- (2,3,0);
\draw[->] (2,2,0) -- (3,3,0);

\node[below right, outer sep=0.5pt, fill = white] (0,0,0) {O};
\end{tikzpicture}

\begin{theorem} 
A stationary distribution exists for the Markov chain and it is unique.
\end{theorem}
\textit{\textbf{Proof:}}\\
\textit{Irreducibility}:
Graphically, we can see that its a random walk on three surfaces of a cube where it is possible to move along the grid and diagonals ($x=y$ line). So, from each point in the graph it is possible to move to any other point in a finite number of steps.\\
Alternatively, we arrange the states in the following manner: \\
Start at origin $(0,0,0)$. Keeping $a_3=0$ fixed, we move on the $X-Y$ plane in the following manner. \\
$(0,0,0) \rightarrow (1,0,0) \rightarrow (2,0,0) \rightarrow...\rightarrow (\kbar,0,0) \rightarrow (\kbar,1,0) \rightarrow (\kbar,2,0) \rightarrow ... \rightarrow (\kbar,\kbar,0) \rightarrow \\
(\kbar-1,\kbar,0) \rightarrow...\rightarrow (1,\kbar,0) \rightarrow (1,\kbar-1,0) \rightarrow (2,\kbar-1,0) \rightarrow...\rightarrow (\kbar-1,\kbar-1,0)\rightarrow...\rightarrow (2,2,0)\rightarrow (1,2,0)\rightarrow (1,1,0)  \rightarrow\\
(0,1,0) \rightarrow ....(0,1,1) \rightarrow (0,0,1) \rightarrow ... \rightarrow (1,0,1)$\\
In this arrangement of states, every state has a positive probability of going to the next stage in both directions. This shows that the Markov Chain is irreducible. \\
\textit{Aperiodicity}:
For each state, there is a positive probability of occurrence of $(L_1,L_2,L_3)$, which means that there is a positive probability of remaining in that state. So the diagonals of the Transition matrix is positive. Hence the Markov Chain is aperiodic. \\
An irreducible aperiodic finite state Markov Chain must be positive recurrent. Hence it is ergodic and a unique stationary distribution exists. 
\subsubsection{Finding the Stationary Distribution}

For ease of understanding we first find the stationary distribution for $\kbar = 2$. The transition matrix then becomes $T_2$, where,
\[
\frac{1}{pq}T_{2} = \begin{bmatrix} 
    \frac{1}{pq}-3 & q \undertilde{\mathbb{1}}^T_3 & p \undertilde{\mathbb{1}}^T_3  & \undertilde{0}^T_3  & \undertilde{0}^T_6 & \undertilde{0}^T_3\\
    p \undertilde{\mathbb{1}}_3 & (\frac{1}{pq}-3)I_3 & qS  & qI_3 & pI_3 \otimes \undertilde{\mathbb{1}}^T_2 & O^{3 \times 3} \\
    q \undertilde{\mathbb{1}}_3 & pS & (\frac{1}{pq}-3)I_3 & O^{3 \times 3} & q\undertilde{\mathbb{1}}^T_2 \otimes  I_3& pI_3\\
    \undertilde{0}_3 & pI_3 & O^{3 \times 3} & (\frac{1}{pq}-p-2q)I_3 & qI_3 \otimes \undertilde{\mathbb{1}}^T_2 & O^{3 \times 3}\\
    \undertilde{0}_6 & qI_3 \otimes \undertilde{\mathbb{1}}_2 & p\undertilde{\mathbb{1}}_2 \otimes I_3 & I_3 \otimes \undertilde{\mathbb{1}}_2 & (\frac{1}{pq}-2q-p-1)I_6  & q\undertilde{\mathbb{1}}_2 \otimes I_3\\
    \undertilde{0}_3 & O^{3 \times 3} & qI_3 & O^{3 \times 3} & \undertilde{\mathbb{1}}_2^T \otimes I_3 & (\frac{1}{pq}-2-q)I_3
    \end{bmatrix}
\]
Now, to find stationary distribution of this Markov chain, we need to solve for $\pi T_2 = \pi$. \\
For $\kbar = 2, \pi = (\pi_1,\pi_2,...,\pi_{19})$ is a vector of length $19$. Moreover, $\sum\limits_{i=1}^{19} \pi_i = 1$. So we have 20 equations in 19 unknowns. 
However, since the proportion of high types is same for all populations, so we can assume that in the long run, the time spent by the Markov Chain on $(a_{i_1},a_{i_2},a_{i_3})$ is same as that spent on $(a_{j_1},a_{j_2},a_{j_3})$, where $(j_1,j_2,j_3)$ is a permutation of $(i_1,i_2,i_3)$.
So the $\pi_i$'s corresponding to these states will be equal.
This reduces the system of equations to a set of 7 equations in 6 unknowns. Let us call them $x_1,x_2,..x_6$. Then the equations simplify to:
\begin{subequations}
    \begin{align}
    x_1 & = px_2 + qx_3\\
    3x_2 & = qx_1 + 2px_3 + px_4 + 2qx_5\\
    3x_3 & = px_1 + 2qx_2 + 2px_5 + + qx_6\\
    (2-p) x_4 & = qx_2 + 2x_5\\
    (3-p) x_5 & = px_2 + qx_3 + qx_4 + x_6\\
    (3-p) x_6 & = px_3 + 2qx_5\\
    1 & = x_1 + 3x_2+ 3x_3+ 3x_4+ 6x_5 + 3x_6
\end{align}
\end{subequations}
Solving the system of equations, we get,
\begin{align*}
    x_1 & = (pA+q)x_3\\
    x_2 & = Ax_3;\\
    & A = \frac{p^5-6p^4+18p^3-34p^2+31p-20}{p^5-6p^4+17p^3-30p^2+28p-18}\\
    x_3 & = \frac{(p^2-2p+2)(p^5 - 6p^4 + 17p^3 - 30p^2 + 28p - 18)}{19p^7 - 163p^6 + 670p^5 - 1707p^4 + 2800p^3 - 3079p^2 + 2086p - 786}\\
    x_5 & = Bx_3;\\
    x_4 & = \frac{qA + 2B}{2-p}x_3\\
    & B = \frac{p^7 - 9p^6 + 38p^5 - 97p^4 + 159p^3 - 173p^2 + 116p - 42}{(p^5 - 6p^4 + 17p^3 - 30p^2 + 28p - 18)(p^2-2p+2)}\\
    x_6 & = \frac{p + 2qB}{3-p}x_3
\end{align*}
See Appendix~\ref{app:proof-thm1} for the detailed proof.

For threshold value 2, Figure 1 gives the stationary probabilities of the unique states according to the changes in value of $p$. 

    \begin{figure}
        \centering
        \includegraphics[scale=1.1]{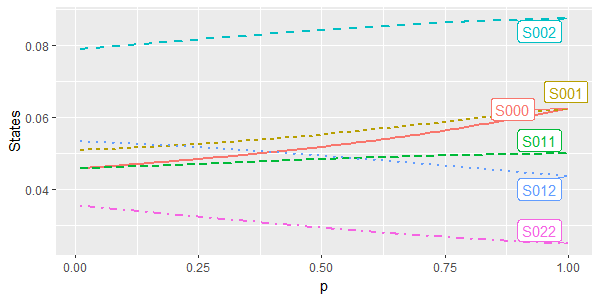}
        \caption{Stationary probabilities of the unique states corresponding to all values of p}
        \label{fig:enter-label}
    \end{figure}

The stationary distribution corresponding to any such $\kbar$ can be solved similarly by solving the $\binom{\kbar}{2}+2\kbar + 2$ equations in $\binom{\kbar}{2}+2\kbar + 1$ unknowns. We have used programming to solve for the stationary distribution given any value of the threshold and any $p$.

Figures 2,3,4 and 5 show the stationary probabilities corresponding to different combinations of the values of $\kbar$ and $p$.

\begin{figure}
    \centering
    \includegraphics[width=0.8\linewidth]{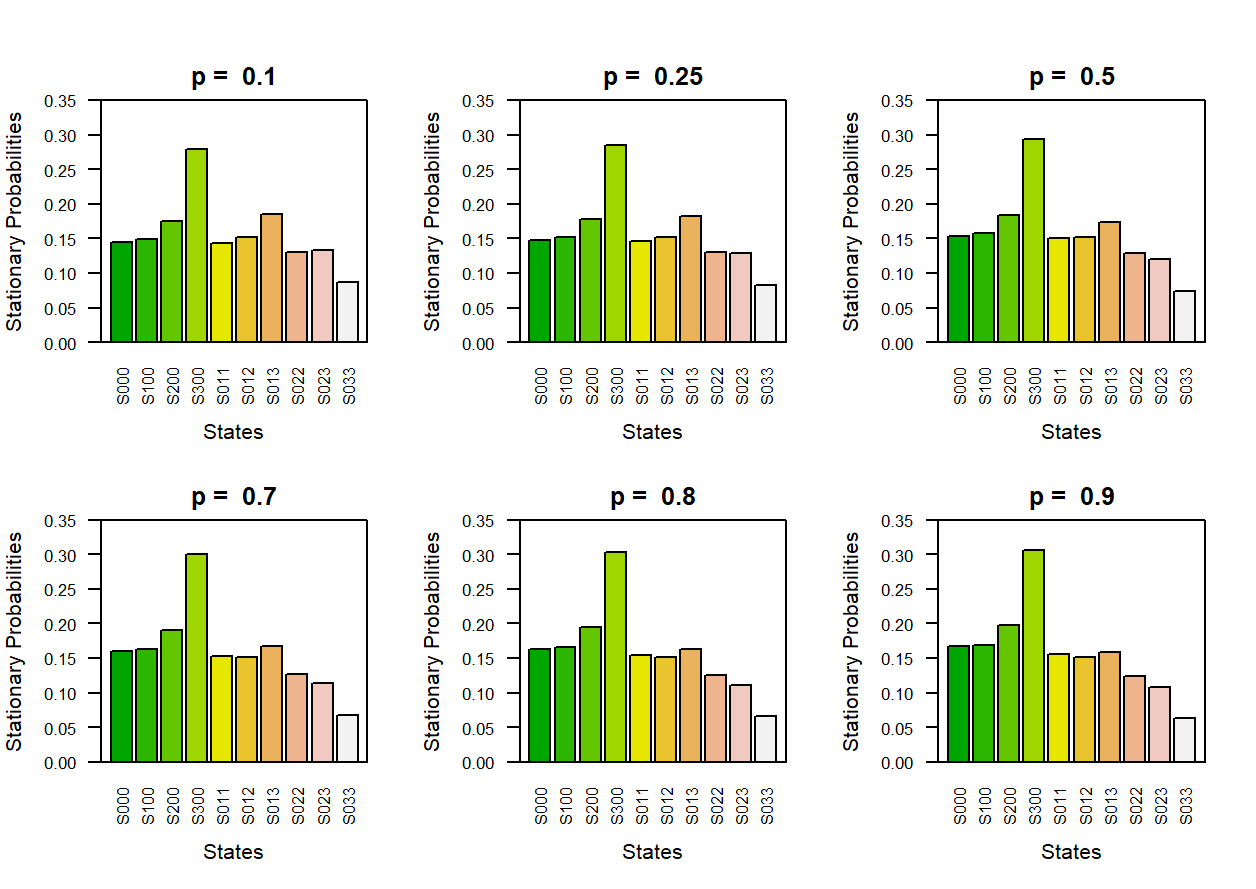}
    \caption{Stationary probabilities of the unique states for different values of p for threshold value = 3}
    \label{fig:enter-label1}
\end{figure}
\begin{figure}
    \centering
    \includegraphics[width=0.8\linewidth]{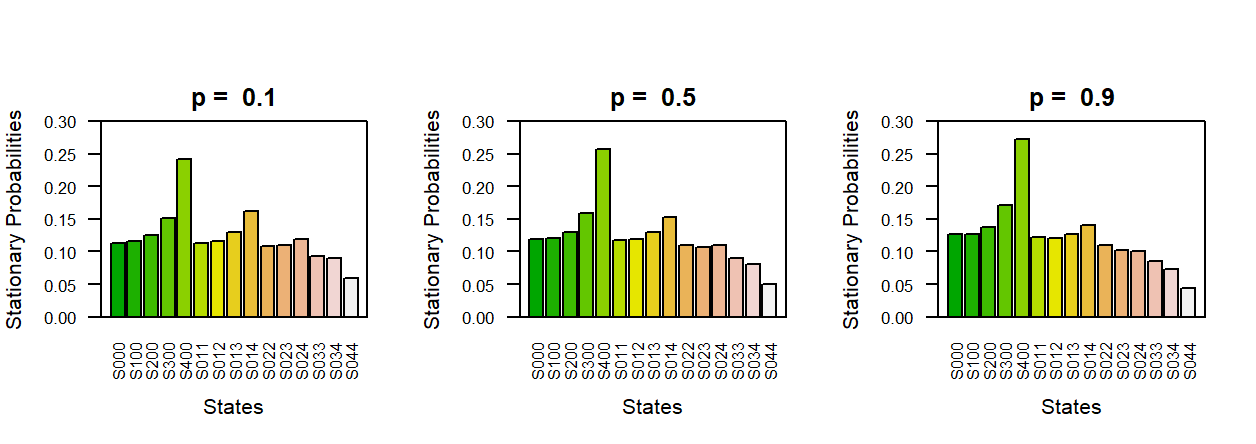}
    \caption{Stationary probabilities of the unique states for different values of p for threshold value = 4}
    \label{fig:enter-label2}
\end{figure}
\begin{figure}
    \centering
    \includegraphics[width=0.8\linewidth]{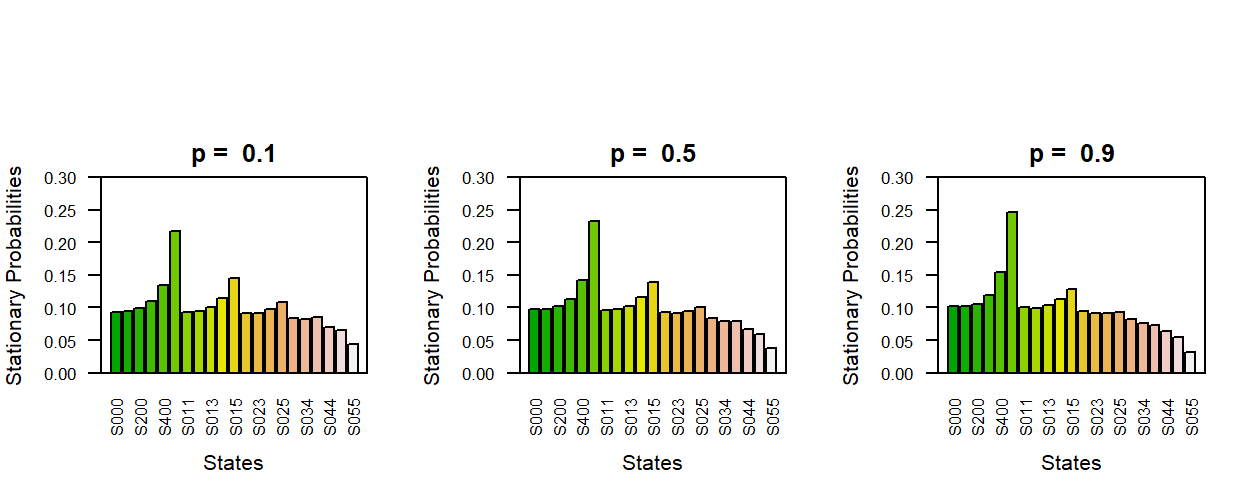}
    \caption{Stationary probabilities of the unique states for different values of p for threshold value = 5}
    \label{fig:enter-label3}
\end{figure}
\begin{figure}
    \centering
    \includegraphics[width=0.8\linewidth]{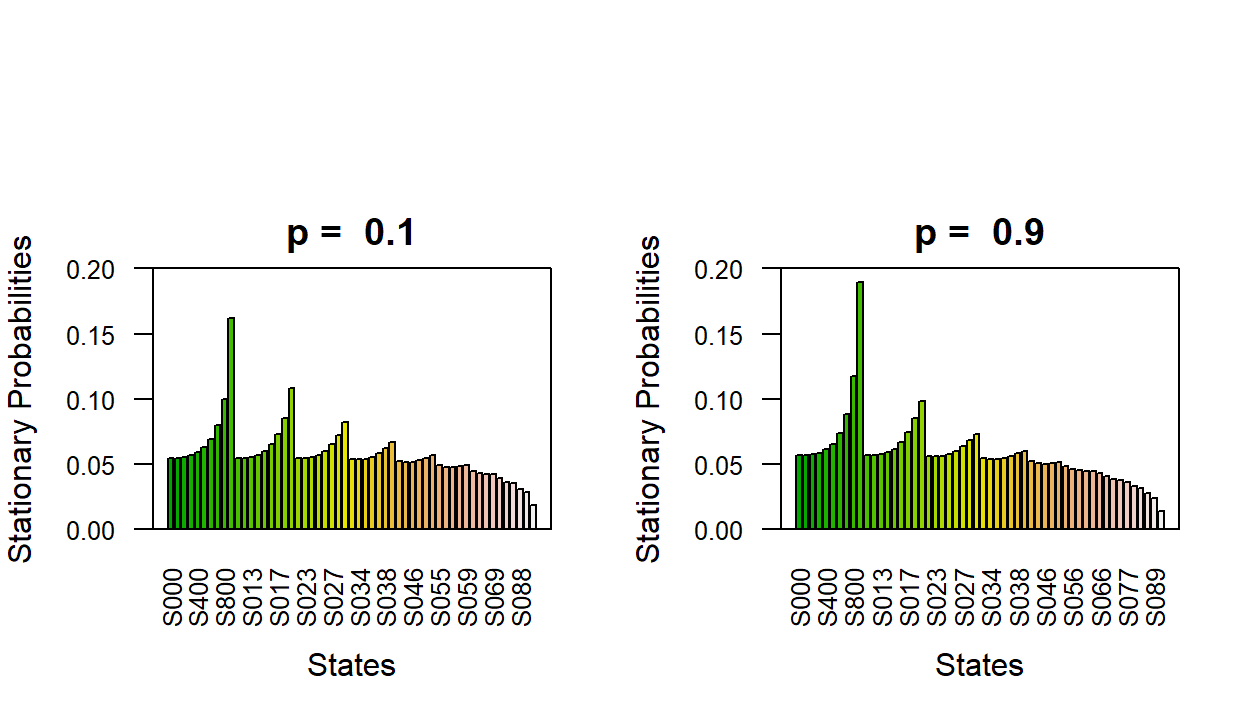}
    \caption{Stationary probabilities of the unique states for different values of p for threshold value = 9}
    \label{fig:enter-label4}
\end{figure}
\newpage
\section{Three Way Dis-assortative Matching}
Here we assume that preference is higher for dissimilar types in a team. The utility for a team with dissimilar types is higher than that with similar type of individuals in them. In the simplest case, 
\begin{align*}{
    U(H_i H_j L_k) = U(H_i L_j L_k) >  U(H_i H_j H_k) , U(L_i L_j L_k) \qquad \forall i\neq j \neq k=1(1)3
    }
\end{align*}
Whenever, there are a mix of high and low types in the market they get matched to form a team. So, the only possibility is either all lows or all high types remaining in the market. There will be same number of either type from each population (since they arrive as a triplet). If the number of high types exceeds $\kbar_H$, excess high types are matched immediately. Similarly if the number of low types crosses $\kbar_L$, excess low types are immediately matched. Let $k^t_H$ denote the number of remaining triplet of high types and $k^t_L$ the number of remaining triplet of low types in the market.
Let $k^t = k^t_H - k^t_L$. Now,
\begin{align*}
    k^{t+1} & = k^{t} + 1 && \text{if $(H_1,H_2,H_3)$ appears}\\
    k^{t+1} & = k^{t} - 1 && \text{if $(L_1,L_2,L_3)$ appears}\\
    k^{t+1} & = k^{t} && \text{if $k^t < 0$ and $(H_i,L_j,L_k)$ appears, $i \neq j \neq k = 1(1)3$}\\
    k^{t+1} & = k^{t} + 1 && \text{if $k^t < 0$ and $(H_i,H_j,L_k)$ appears, $i \neq j \neq k = 1(1)3$}\\
     k^{t+1} & = k^{t} && \text{if $k^t > 0$ and $(H_i,H_j,L_k)$ appears, $i \neq j \neq k = 1(1)3$}\\
    k^{t+1} & = k^{t} - 1 && \text{if $k^t > 0$ and $(H_i,L_j,L_k)$ appears, $i \neq j \neq k = 1(1)3$}
\end{align*}
This process induces a Markov chain in the following manner. Let $
x^t_i = I_{[k^{t} = i]}$. 
Then, $
x^{t+1} = T_{\overline{k}} x^t$, 
where $T_{\overline{k}}$ is the transition matrix given by

\[
T_{\overline{k}} = \begin{psmallmatrix} 
    q^3 + 3pq^2 & p^3 + 3p^2q & 0 & \dots & 0 & 0 & 0 & \dots & 0 & 0 & 0\\
    q^3 & 3pq^2 & p^3 + 3p^2q & \dots & 0 & 0 & 0 & \dots & 0 & 0 & 0\\
    \vdots & \vdots & \vdots & \vdots & \vdots & \vdots & \vdots & \vdots & \vdots & \vdots & \vdots\\
    0 & 0 & 0 &\dots  & q^3 & 3pq^2 & p^3 + 3p^2q & \dots & 0 & 0 & 0\\
    0 & 0 & 0 & \dots  & q^3 & 1-p^3-q^3 & p^3 & \dots & 0 & 0 & 0\\
    0 & 0 & 0 & \dots  & 3pq^2+q^3 & 3p^2q & p^3 & \dots & 0 & 0 & 0\\
    \vdots & \vdots & \vdots & \vdots & \vdots & \vdots & \vdots & \vdots & \vdots & \vdots & \vdots\\
    0 & 0 & 0 & \dots & 0 & 0 & 0 & \dots & 3pq^2+q^3 & 3p^2q & p^3\\
    0 & 0 & 0 & \dots & 0 & 0 & 0 & \dots & 0 & q^3+3pq^2 & p^3+3p^2q\\
    \end{psmallmatrix}
\]
 It can be shown that this Markov chain is aperiodic, irreducible and positively recurrent. So it is an ergodic Markov chain. Hence, an unique stationary distribution exists. Solving for the stationary distribution we get,
\begin{align*}
     \pi_i & = \frac{q^3}{p^3+3p^2q} \pi_{i+1} && \qquad i=-\kbar_L,-\kbar_L+1,...,-1\\
     \pi_i & = \frac{3pq^2 + q^3}{p^3} \pi_{i+1} && \qquad i=0,1,...,\kbar_H-1
\end{align*}
Or,
\begin{align*}
     \pi_i & = a \pi_{i+1};  && a = \frac{q^3}{p^3+3p^2q} &&& \qquad i=-\kbar_L,-\kbar_L+1,...,-1\\
     \pi_{i+1} & = b \pi_{i}; && b = \frac{p^3}{3pq^2 + q^3} &&& \qquad i=0,1,...,\kbar_H-1
\end{align*}
Here,
\begin{align*}
    \pi_0 = \frac{1}{\left[ 1 + \frac{a(1-a^{\kbar_L})}{1-a} + \frac{b(1-b^{\kbar_H})}{1-b} \right]}
\end{align*}
The stationary distribution is a mixture of two truncated geometric distributions. 

Figures 6,7 and 8 shows the stationary probabilities corresponding to each state taking different combination of values of $p$, $\kbar_L$ and $\kbar_H$. The names High and Low in the title of the figures indicate the thresholds corresponding to the High and the Low Types respectively. 
\begin{figure}[H]
    \centering
    \includegraphics[width=0.82\linewidth]{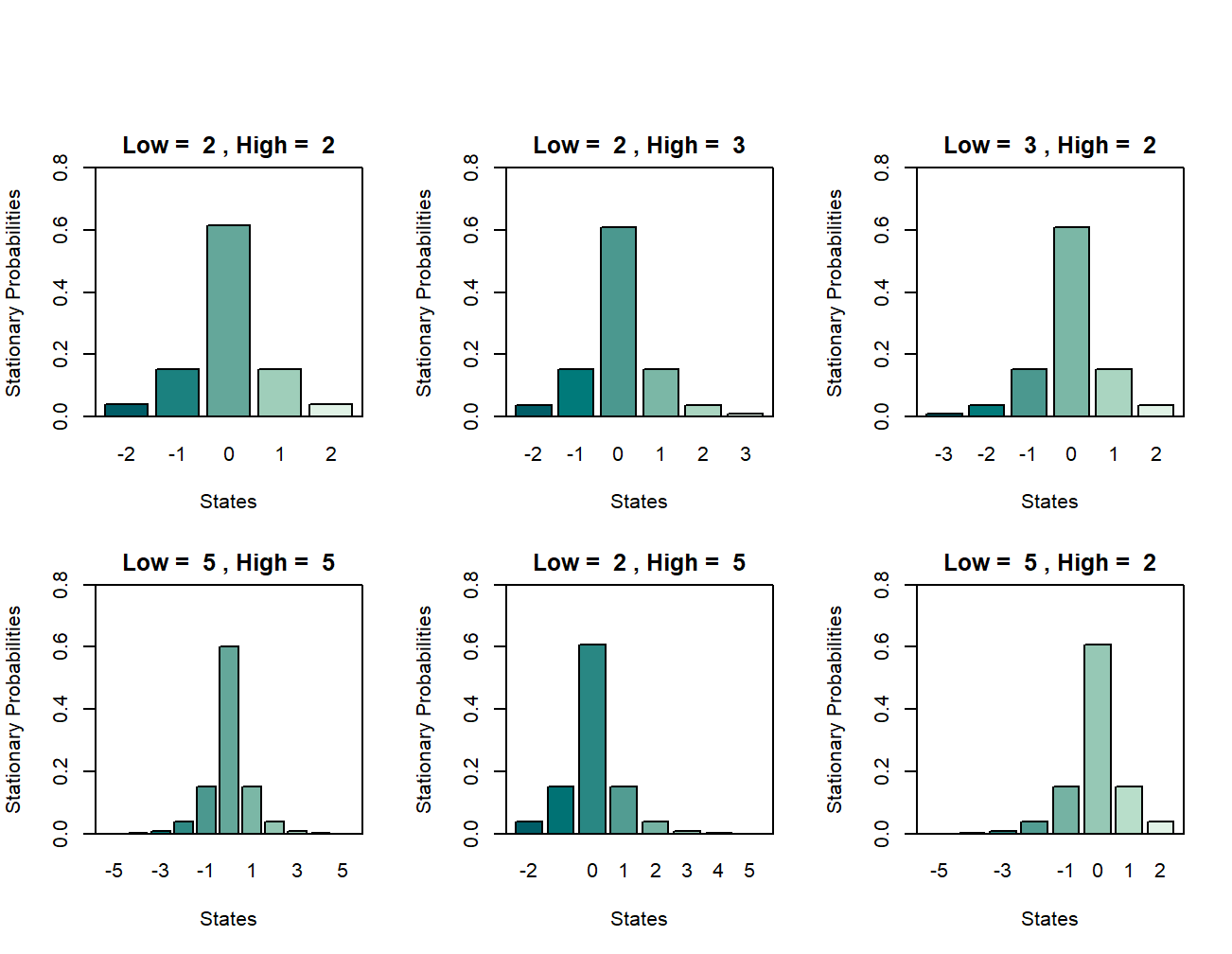}
    \caption{Stationary probabilities of the states for different values of thresholds corresponding to p = 0.5}
    \label{fig:enter-label5}
\end{figure}
\vspace{-1em}
\begin{figure}[H]
    \centering
    \includegraphics[width=0.82\linewidth]{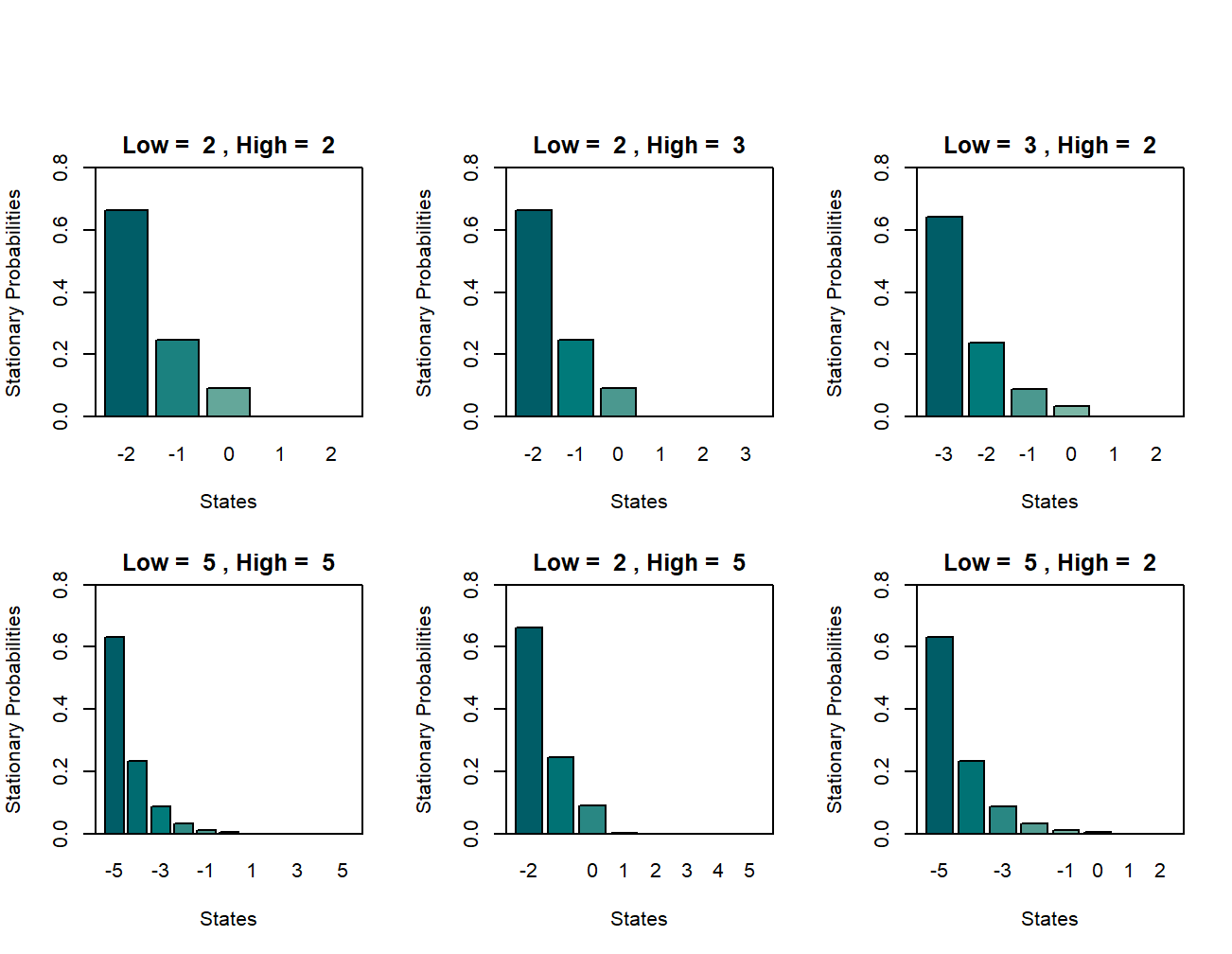}
    \caption{Stationary probabilities of the states for different values of thresholds corresponding to p = 0.25}
    \label{fig:enter-label6}
\end{figure}
\begin{figure}[H]
    \centering
    \includegraphics[width=0.82\linewidth]{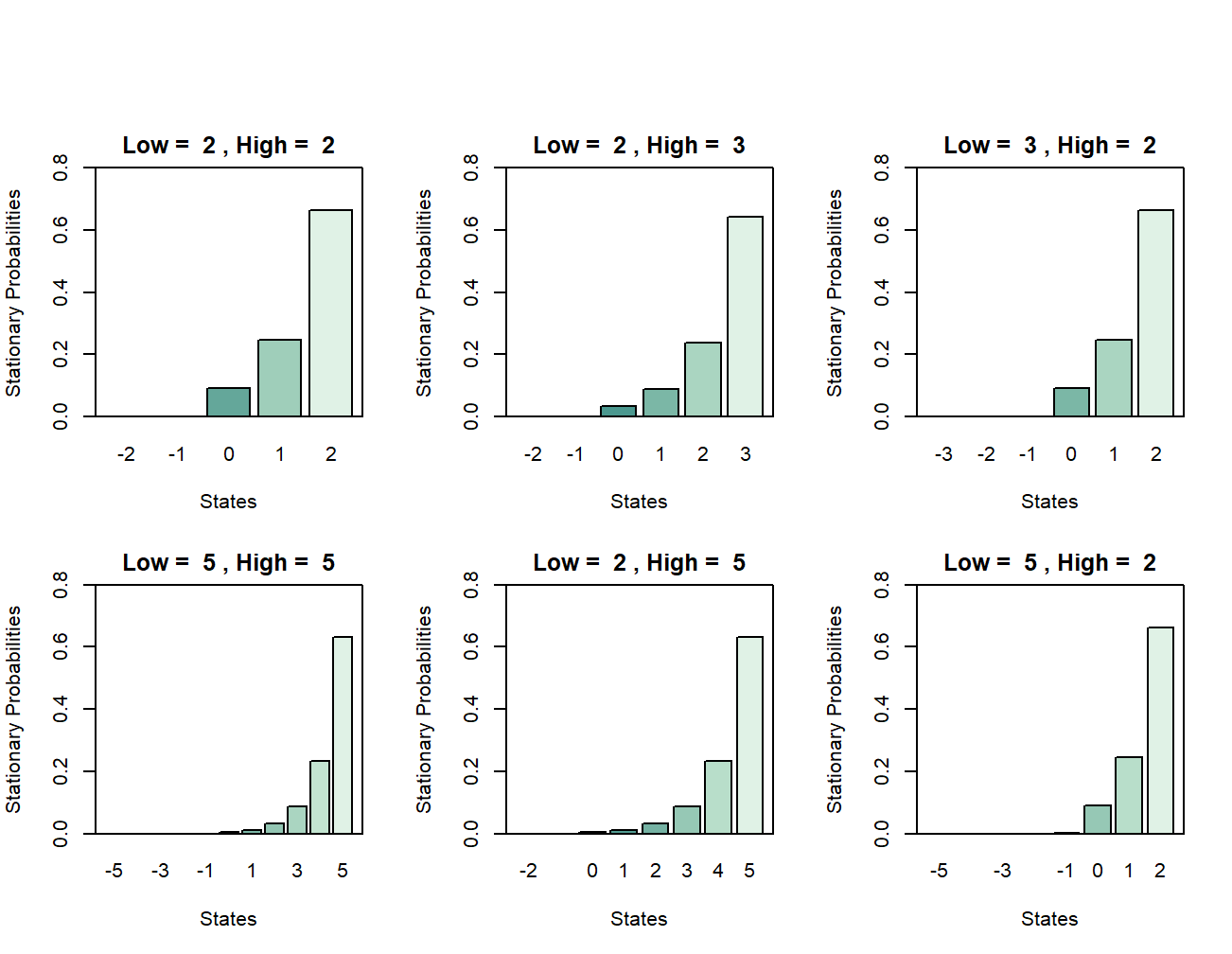}
    \caption{Stationary probabilities of the states for different values of thresholds corresponding to p = 0.75}
    \label{fig:enter-label7}
\end{figure}

	\bibliographystyle{plainnat}
	\setcitestyle{numbers}
	\bibliography{mybib}
\appendix
\begin{appendices}
\section{Proof of Stationary Distribution of Three-Way Assortative Matching for Threshold Value = 2}\label{app:proof-thm1}

The transition probability matrix corresponding to $\kbar=2$ is $T_2$, where,
\[
\frac{1}{pq}T_{2} = \begin{bmatrix} 
    \frac{1}{pq}-3 & q \undertilde{\mathbb{1}}^T_3 & p \undertilde{\mathbb{1}}^T_3  & \undertilde{0}^T_3  & \undertilde{0}^T_6 & \undertilde{0}^T_3\\
    p \undertilde{\mathbb{1}}_3 & (\frac{1}{pq}-3)I_3 & qS  & qI_3 & pI_3 \otimes \undertilde{\mathbb{1}}^T_2 & O^{3 \times 3} \\
    q \undertilde{\mathbb{1}}_3 & pS & (\frac{1}{pq}-3)I_3 & O^{3 \times 3} & q\undertilde{\mathbb{1}}^T_2 \otimes  I_3& pI_3\\
    \undertilde{0}_3 & pI_3 & O^{3 \times 3} & (\frac{1}{pq}-p-2q)I_3 & qI_3 \otimes \undertilde{\mathbb{1}}^T_2 & O^{3 \times 3}\\
    \undertilde{0}_6 & qI_3 \otimes \undertilde{\mathbb{1}}_2 & p\undertilde{\mathbb{1}}_2 \otimes I_3 & I_3 \otimes \undertilde{\mathbb{1}}_2 & (\frac{1}{pq}-2q-p-1)I_6  & q\undertilde{\mathbb{1}}_2 \otimes I_3\\
    \undertilde{0}_3 & O^{3 \times 3} & qI_3 & O^{3 \times 3} & \undertilde{\mathbb{1}}_2^T \otimes I_3 & (\frac{1}{pq}-2-q)I_3
    \end{bmatrix}
\]
Now, to find stationary distribution of this Markov chain, we need to solve for $\pi T_2 = \pi$. \\
For $\kbar = 2, \pi = (\pi_1,\pi_2,...,\pi_{19})$ is a vector of length $19$. Moreover, $\sum\limits_{i=1}^{19} \pi_i = 1$. So we have 20 equations in 19 unknowns. 
However, since the proportion of high types is same for all populations, so we can assume that in the long run, the time spent by the Markov Chain on $(a_{i_1},a_{i_2},a_{i_3})$ is same as that spent on $(a_{j_1},a_{j_2},a_{j_3})$, where $(j_1,j_2,j_3)$ is a permutation of $(i_1,i_2,i_3)$.
So the $\pi_i$'s corresponding to these states will be equal.
This reduces the system of equations to a set of 7 equations in 6 unknowns. Let us call them $x_1,x_2,..x_6$. Then the equations simplify to:
\begin{subequations}
    \begin{align}
    x_1 & = px_2 + qx_3\\
    3x_2 & = qx_1 + 2px_3 + px_4 + 2qx_5\\
    3x_3 & = px_1 + 2qx_2 + 2px_5 + + qx_6\\
    (2-p) x_4 & = qx_2 + 2x_5\\
    (3-p) x_5 & = px_2 + qx_3 + qx_4 + x_6\\
    (3-p) x_6 & = px_3 + 2qx_5\\
    1 & = x_1 + 3x_2+ 3x_3+ 3x_4+ 6x_5 + 3x_6
\end{align}
\end{subequations}
Putting the value of $x_4, x_6$ from Eqs. (1d) and (1f) in Eq. (1e) and the value of $x_1,x_4$ from Eqs. (1a) and (1d) in Eq. (1b)
\begin{subequations}
\begin{align}
    (8-7p+4p^2-p^3) x_5 & = (3-p)x_2 + (p^2-3p+3)(2-p)x_3\\
    -2(p^2 - 2p + 2)x_5 & = (2-p)(p^2+1)x_3 + (p^3 - 4p^2+6p-6)x_2
\end{align}
\end{subequations}
Equating the value of $x_5$,
\begin{align*}
    x_2 & = Ax_3;
    & \qquad A = \frac{p^5-6p^4+18p^3-34p^2+31p-20}{p^5-6p^4+17p^3-30p^2+28p-18}
\end{align*}
Putting this in (2b)
\begin{align*}
    x_5 & = Bx_3;
    & \qquad B = \frac{p^7 - 9p^6 + 38p^5 - 97p^4 + 159p^3 - 173p^2 + 116p - 42}{(p^5 - 6p^4 + 17p^3 - 30p^2 + 28p - 18)(p^2-2p+2)}
\end{align*}
\begin{align*}
    x_1 & = (pA+q)x_3\\
    x_4 & = \frac{qA + 2B}{2-p}x_3\\
    x_6 & = \frac{p + 2qB}{3-p}x_3;\\
   1& = \left(pA+\left(1-p\right)+3A+3+\frac{3\left(qA+2B\right)}{2-p}+6B+\frac{3(p+2qB)}{3-p}\right)x_3\\
   XA+YB+Z & = 1
\end{align*}
\begin{align*}
   X & = \frac{-p^2-4p+9}{2-p}A\\
   Y & = \frac{12p^2-54p+66}{(2-p)(3-p)}B\\
   Z & = \frac{p^2-4p+12}{3-p}\\
   \text{Solving,}\\
   x_3 & = \frac{(p^2-2p+2)(p^5 - 6p^4 + 17p^3 - 30p^2 + 28p - 18)}{19p^7 - 163p^6 + 670p^5 - 1707p^4 + 2800p^3 - 3079p^2 + 2086p - 786}
\end{align*} 
\end{appendices}

\end{document}